\documentclass[aps,prb,reprint,amsmath,amssymb,groupedaddress]{revtex4-2}
\usepackage{bm}
\usepackage{graphics}
\usepackage{graphicx}
\usepackage{color}
\usepackage{hyperref}
\usepackage[mathlines]{lineno}

\begin{document}

%\preprint{}


\title{Twisting or untwisting graphene twisted nanoribbons without
  rotation}


\author{Alexandre F. Fonseca}
\email[]{afonseca@ifi.unicamp.br}
%\homepage[aff]{https://www.ifi.unicamp.br/~afonseca/}
%\thanks{}
\affiliation{Applied Physics Department, Institute of Physics ``Gleb
  Wataghin'', State University of Campinas, Campinas, SP, 13083-970,
  Brazil.}


\date{\today}

\begin{abstract}
The common sense regarding twisting or untwisting a ribbon is that it
requires the application of an external rotation to happen. However,
at nanoscale, the application of precise amounts of rotation on a
nanoribbon is not a trivial task. Here, the concept of an alternative
method to add twist to or remove twist from a twisted graphene
nanoribbon (TGNR) without rotation is presented. The method consists
of suspending a TGNR on two separate substrates and by changing only
their distance, the total amount of twist of the TGNR is shown to
change. The possibility to fine-tuning the amount of twist of a TGNR
is also shown. The concept is demonstrated through fully atomistic
molecular dynamics simulations and numerical calculations of the
topological parameters {\it twist} and {\it writhe} of a TGNR. It is
shown that the above process satisfies the so-called {\it linking
  number} theorem of space curves. Besides being experimentally
feasible, this concept reveals a new kind of {\textit{twist to writhe
    transition}} phenomenon that is tension-free and does not require
controlling neither the nanoribbon end-to-end distance nor its
critical twist density.
    \end{abstract}

%\keywords{}

\maketitle

\section{Introduction}

Graphene nanoribbons (GNRs) have the most of the outstanding
properties of pristine graphene and a non-null bandgap that is
edge-dependent and inversely proportional to the ribbon
width~\cite{1,2,3}. GNRs have been fabricated by several methods, from
top-down to chemical and bottom-up
methods~\cite{4,5,6,7,8,10,11,12,13}. A particularly interesting
discovery is the possibility of tuning the bandgap and other
electronic and magnetic properties of GNRs by application of twisting
along its axis~\cite{14,15,16,17,18,19,20,22}. Electronic,
mechanical~\cite{24,25,26,27,28} and thermal~\cite{29,30,31}
properties of twisted GNRs (TGNRs) make them promising and versatile
nanostructures for diverse applications.

Fabrication of TGNRs has been reported in the literature. Chamberlain
{\it et al.}~\cite{33} grew TGNRs inside carbon nanotubes from
reaction of small sulfur-containing molecules. Cao {\it et
  al.}~\cite{34} developed a method to curl GNRs by thermal annealing
that was used by Jarrari {\it et al.}~\cite{35} to show that curled
GNRs enhance photocurrent responses. Previously developed methods to
bend and twist nanotubes~\cite{36} or induce, by laser, changes in
GNRs~\cite{37}, might be useful to fabricate TGNRs.

A ubiquitous phenomenon in filamentary structures is the so-called
{\it twist-to-writhe transition} (TWT). It consists of releasing the
torsional elastic energy accumulated in an initially straight twisted
rod by spontaneous curling and coiling. The TWT is shown to happen
when the filament twist density reaches a critical
value~\cite{40,goriely2000,fonseca2006JAP,mahadevan2019}. In turn, the
twist density is shown to depend on either the applied tensile stress
or filament end-to-end
distance~\cite{goriely2000,fonseca2006JAP,mahadevan2019}. TWT has been
shown to obey the conservation of a geometric quantity called the {\it
  linking number}, $Lk$, of a pair of closed curves or a pair of open
curves with extremities prevented from crossing one with respect to
the other. Defined as a double Gauss integral along the two curves,
$Lk$ is shown to be always an integer number given by half the value
of a certain ``oriented'' way of counting how many times one curve
crosses the other~\cite{writhe2006}. $Lk$, then, satisfies the
C\u{a}lug\u{a}reanu-White-Fuller {\it linking number}
theorem~\cite{40,39,41}:
\begin{equation}
  \label{cwf}
  Lk = Tw + Wr \, ,
\end{equation}
where $Tw$ ($Wr$) is the {\it total twist} ({\it writhe}) of the
filament (filament centerline). {\it Writhe} is a geometric quantity
related to the non-planarity of a space
curve~\cite{41,writhe2006,kleninwrithe2000}.

TWT is observed in conformations of
DNA~\cite{kleninwrithe2000,dna0,dna1,dna2,dna3}, filamentary
structures of some bacteria~\cite{mendel1,mendel2,prl1998}, in garden
hoses, telephone cords, cables and other engineering
structures~\cite{dna3,c1,c2,c3}. It is also present in a wide range of
correlated phenomena and applications as in dynamics of stiff
polymers~\cite{prl2000}, coiled artificial muscles made of twisted
carbon nanotube yarns~\cite{science2012} or fishing
lines~\cite{science2014}, helicity in solar active
structures~\cite{magnetic2014} and in fluid
vortices~\cite{science2017vortex}, chemical synthesis of twisted
annulene molecules~\cite{natchem2014annulene}, mechanics of knots and
tangles~\cite{science2020}, collagen
fibrils~\cite{acsnano2020collagen}, etc.

In this Work, a new experimental concept designed to promote and
control the interconversion between {\it twist} and {\it writhe} in a
TGNR, without rotation of its extremities, is proposed and
computationally demonstrated. Basically, it consists of laying the
extremities of a TGNR on two separate substrates, and then allowing
the distance between them to vary, within the nanoribbon length
(Fig. \ref{fig1}). As these substrates play an essential role on the
proposed TWT phenomenon, it is here named {\it substrate induced} TWT
(SITWT). Although nanoribbons can be subject to regular
TWT~\cite{twcTGNR2014scirep}, the proposed interconversion method is
innovative because it does not require the TGNR to be neither
tensioned (or tension released) nor additionally twisted to reach or
exceed the critical twist density.

Section \ref{sec2} presents the description of the proposed SITWT
method as well as 
%of changing the total twist of a TGNR without applying rotation to
%its ends. It also presents
the theory and the computational approach used to calculate $Wr$ and
$Tw$ of each configuration of the TGNR, required to demonstrate the
SITWT. It also describes the computational methods employed to
simulate the SITWT experiment. In Sections \ref{sec3} and \ref{sec4},
the results and the conclusions are presented, respectively.

\section{Methodology}
\label{sec2}

\subsection{Description of the proposed SITWT method}

\begin{figure}[h!]
 \includegraphics[width=8.5cm,keepaspectratio]{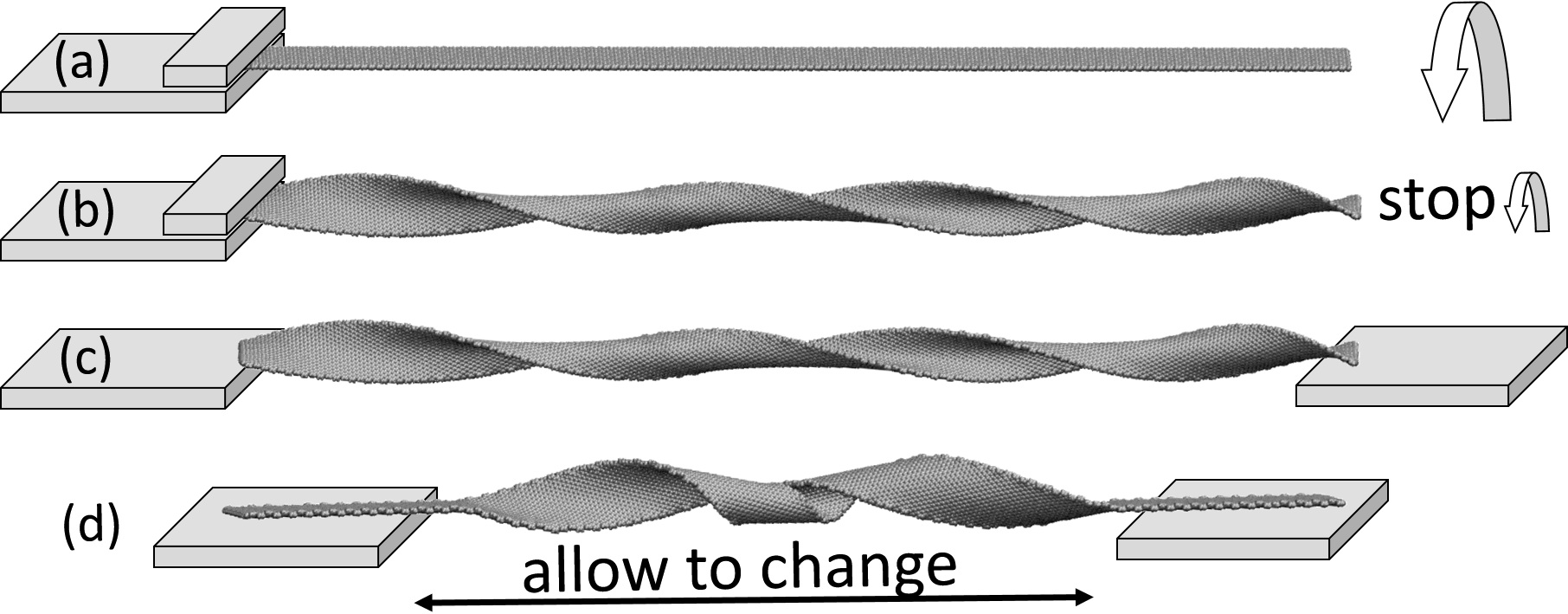}
 \caption{\label{fig1} Experimental scheme to demonstrate the SITWT.
   Panels (a) and (b) show the initial preparation of a TGNR by fixing
   one extremity of the straight untwisted GNR (a), then applying a
   torsional strain until reaching the desired amount of initial total
   twist (b). Panels (c) and (d) show both TGNR extremities being laid
   on two substrates without additional constraints, and the distance
   between the substrates being allowed to change.}
\end{figure}

%The description of the proposed SITWT experiment is given as follows.
An initial amount of turns or torsional strain has to be applied to a
GNR in order to produce a TGNR with a given value of $Lk$
(Figs. \ref{fig1}a and \ref{fig1}b). $Lk$ will be conserved as long as
the TGNR extremities are prevented from rotation with respect to the
ribbon axis (also called the ribbon centerline).

The experiment itself consists of first suspending the TGNR, by laying
its two extremities on two separated substrates (Fig. \ref{fig1}c). As
the adhesion forces between the nanoribbon and substrates are
relatively large, as in graphene-graphene surface
interactions~\cite{annett2016nature,fonseca2019Carbon}, it is expected
that these forces will themselves prevent the TGNR extremities from
rotating or releasing the initial applied torsional strain. The idea
of the proposed experiment is, then, to allow the distance between the
substrates to vary within the size of the TGNR (Fig. \ref{fig1}d).
Variation of this distance leads to variation of the amount of TGNR
surface that interacts with the substrates. As the flexural rigidity
of nanoribbons are usually low (see for example, that of
graphene~\cite{zerobending2011prl}), van der Waals forces between the
nanoribbon and substrates flattens the TGNR parts in touch with the
substrates, leading to an overall change of the shape of the suspended
part of the TGNR (illustrated in Fig. \ref{fig1}d). Smaller the
distance between the substrates, larger the difference in the
conformation of the axis (or centerline) of the TGNR from that of a
straight twisted ribbon. As a consequence, the {\it writhe}, $Wr$, of
the TGNR centerline changes with the substrate distance. As long as
the adhesion forces with the substrates keep preventing the TGNR ends
from rotation, the {\it linking number} theorem, eq. (\ref{cwf}), is
expected to be satisfied during the movement of the substrates. The
theorem, then, predicts that if the {\it writhe}, $Wr$, of the TGNR
varies, its {\it total twist}, $Tw$, will vary in order to keep the
TGNR $Lk$ unchanged. That is the basis for the experiment of changing
the twist, $Tw$, without applying or removing any amount of rotation
to the TGNR extremities.

In order to demonstrate the above SITWT, fully atomistic classical
molecular dynamics (MD) simulations of a TGNR suspended on two
substrates will be performed. The AIREBO potential~\cite{old39,old40}
and LAMMPS package~\cite{old41} will be employed to simulate the
proposed experiment of moving substrates with suspended TGNRs.
Graphite substrates will be considered and modeled as fixed graphene
layers. AIREBO is a well-known reactive empirical potential, largely
used to study the structure, mechanical and thermal properties of
carbon
nanostructures~\cite{42,43,44,45,46,47,48,49,50,51,52}. Therefore, the
MD results for the structure and dynamics of the TGNRs on the moving
substrates are expected to really represent real experiments.

From the MD results, the {\it linking number} theorem will be shown to
be always satisfied for suspended TGNRs under the present method. To
show that, the calculation of $Tw$ and $Wr$ for every configuration of
the TGNRs studied here is required. Their summation should be equal to
the initially applied $Lk$ to the TGNR, according to eq. (\ref{cwf}).
In turn, calculations of $Tw$ and $Wr$ require the definition of two
space curves corresponding to the TGNR centerline and an adjacent
line. As described ahead, these space curves will be discretized based
on the positions of two sets of carbon atoms along the TGNR, one at
the middle part of the nanoribbon, and the other at about one graphene
lattice of distance from the first, on the side, respectively. A piece
of the TGNR showing these two sets of atoms is shown in the insets of
Fig. \ref{fig2}. In what follows, the details about how these
quantities are calculated and the definitions of $Tw$ and $Wr$ are
presented.

\subsection{Numerical approach for calculating $Wr$ and $Tw$ of TGNRs}
\label{discretizar}

Let vectors $\bm{\mbox{x}}$ and $\bm{\mbox{y}}$ be identified with the
TGNR centerline and an adjacent line bounded to it, as ilustrated by
red and black atoms drawn in the insets of Fig. \ref{fig2},
respectively. $Tw$ and $Wr$ can be calculated by~\cite{41,writhe2006}:
\begin{equation}
  \label{tw}
  Tw = \frac{1}{2\pi}\oint \bm{\mbox{t}}_{\bm{\mbox{x}}(s)}\cdot
  \left( \bm{\mbox{u}}\times\frac{\mbox{d}\bm{\mbox{u}}}{\mbox{d}s}
  \right) \,\mbox{d}s \, ,
\end{equation}
where $s$ and $\bm{\mbox{t}}$ are the arclength and tangent vector of
the TGNR centerline curve $\bm{\mbox{x}}$, respectively, and
$\bm{\mbox{u}}$ is a unit vector orthogonal to $\bm{\mbox{t}}$, and
pointing from $\bm{\mbox{x}}$ to $\bm{\mbox{y}}$. And
\begin{equation}
  \label{wr}
  Wr = \frac{1}{4\pi}\oint_{\bm{\mbox{x}}}\oint_{\bm{\mbox{x}}}
  \frac{(\bm{\mbox{t}}_{\bm{\mbox{x}}(s)}\times\bm{\mbox{t}}_{\bm{\mbox{x}}(s')})
    \cdot (\bm{\mbox{x}}(s)-\bm{\mbox{x}}(s'))}
       {|\bm{\mbox{x}}(s)-\bm{\mbox{x}}(s')|^3} \mbox{d}s\,\mbox{d}s'
       \, .
\end{equation}
While $Lk$ is shown to be always an integer number, $Tw$ and $Wr$ are
real numbers that, for closed or end-constrained rods, can varies as
long as eq. (\ref{cwf}) is satisfied. Eqs. (\ref{tw}) and (\ref{wr})
are defined for closed curves. However, it has been
shown~\cite{writhe2006,vanderheidge2003} that if the tangents at the
endpoints of a finite open centerline are coplanar, an imagined
coplanar closing curve would contribute with zero to $Wr$. Similarly,
it is possible to think of closing curves for the centerline,
$\bm{\mbox{x}}$, and its adjancent line, $\bm{\mbox{y}}$, that do not
cross one to each other, so contributing with zero to the calculation
of $Tw$. In the proposed experiment, the TGNRs are not closed ribbons
but the substrates on which its extremities are laid, are coplanar.

All above quantities are discretized according to the following
definitions:

\begin{subequations}
\label{discreti}
\begin{eqnarray}
  &&\bm{\mbox{x}}=\{\bm{\mbox{x}}_{1},\bm{\mbox{x}}_{2},\ldots,\bm{\mbox{x}}_{i},
    \ldots,\bm{\mbox{x}}_{N-1},\bm{\mbox{x}}_{N}\} \, , \label{x} \\
  &&\bm{\mbox{y}}=\{\bm{\mbox{y}}_{1},\bm{\mbox{y}}_{2},\ldots,\bm{\mbox{y}}_{i},
    \ldots,\bm{\mbox{y}}_{N-1},\bm{\mbox{y}}_{N}\} \, , \label{y} \\
  &&s_{1}=0\quad\mbox{and}\quad s_{i>1}=
  \sum^{i}_{k=2}|\bm{\mbox{x}}_{k}-\bm{\mbox{x}}_{k-1}|\, , \label{s} \\
  &&\mbox{d}s_1=0\quad\mbox{and}\quad \mbox{d}s_{i>1}=s_i-s_{i-1} \, \label{ds} \\
  &&\bm{\mbox{t}}_{1}=0\quad\mbox{and}\quad\bm{\mbox{t}}_{i>1}=
  \frac{\bm{\mbox{x}}_{i}-\bm{\mbox{x}}_{i-1}}
       {|\bm{\mbox{x}}_{i}-\bm{\mbox{x}}_{i-1}|} \, , \label{t} \\
    &&\bm{\mbox{u}}_i=\frac{\bm{\mbox{y}}_i-\bm{\mbox{x}}_i}
          {|\bm{\mbox{y}}_i-\bm{\mbox{x}}_i|} \, , \label{u} \\
          &&\mbox{d}\bm{\mbox{u}}_i\equiv
          \left.\frac{\mbox{d}\bm{\mbox{u}}}{\mbox{d}s}\right|_{i} ,
          \quad\mbox{d}\bm{\mbox{u}}_1=0\quad\mbox{and} \nonumber \\
          &&\mbox{d}\bm{\mbox{u}}_{i>1}=
          \frac{\bm{\mbox{u}}_{i}-\bm{\mbox{u}}_{i-1}}
               {\mbox{d}s_i} \, , \label{du}
\end{eqnarray}
\end{subequations}
where $\bm{\mbox{x}}_{i}$ ($\bm{\mbox{y}}_{i}$) and $N$ are the
position of the $i$-esim atom along the centerline (adjacent line) and
the number of atoms along the centerline of the TGNR, respectively.
For the TGNR studied here, $N = 285$. In eqs. (\ref{discreti}) the
indices go from 1 to $N$.

\subsection{Molecular Dynamics simulations and chosen structures of TGNRs}

\begin{figure}[h!]
 \includegraphics[width=8.5cm,keepaspectratio]{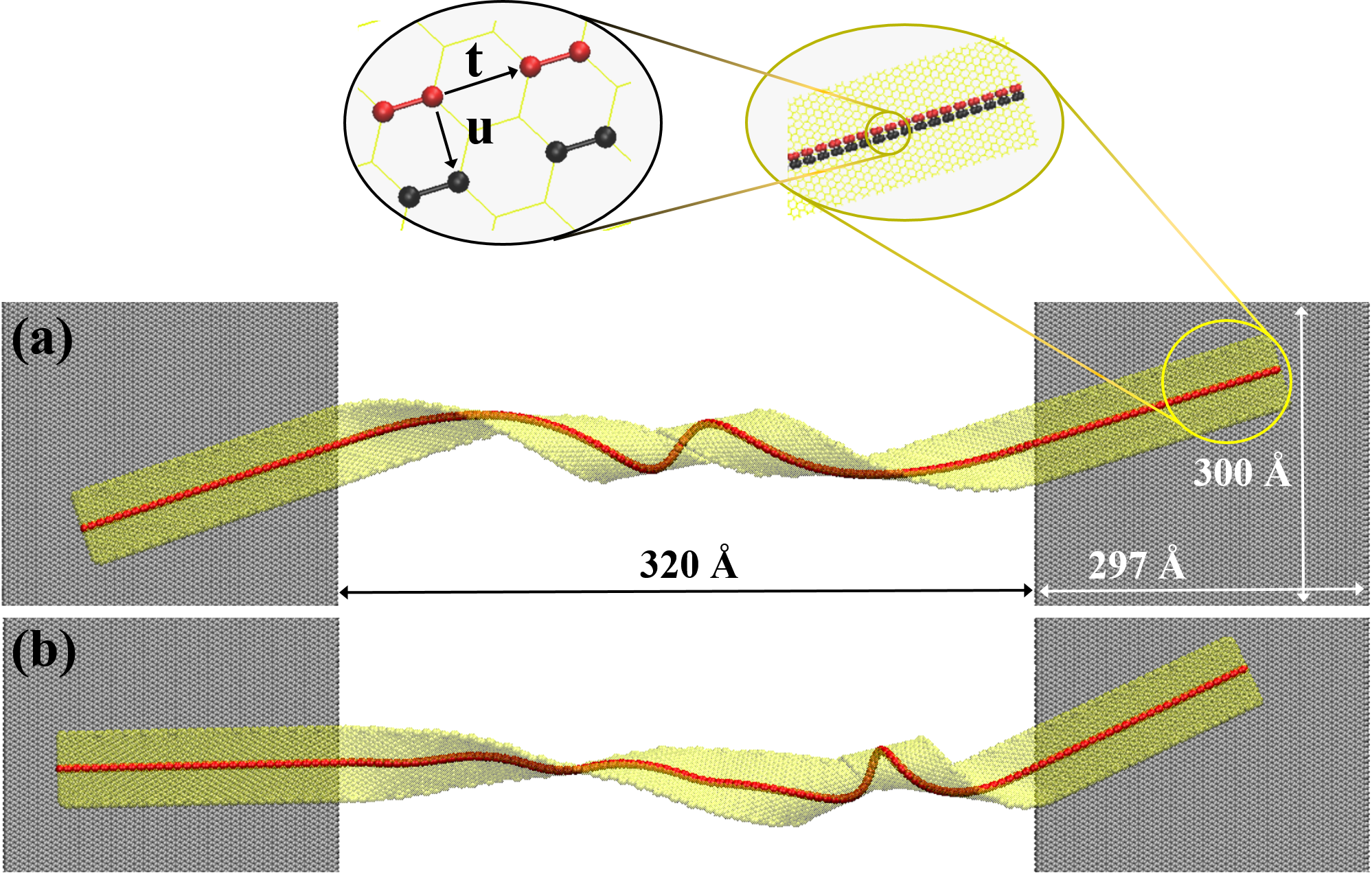}
 \caption{\label{fig2} Upper views of the TGNRs used in the simulated
   SITWT processes. Two substrates of 297 \AA\mbox{} by 300 \AA\mbox{}
   of size (not shown to scale to save space) are placed at an initial
   distance of about 320 \AA. Suspended on these two substrates are
   armchair TGNRs with $Lk=2$ and 600 (33) \AA\mbox{} length
   (width). Panels (a) and (b) show optimized TGNRs after 8 ns of MD
   simulations at 300 K and 1000 K, respectively. Insets show pieces
   of two sets of carbon atoms that represent the centerline (red) and
   an adjacent line (black) of the TGNR. The positions of these sets
   of atoms are used to define and discretize the vectors
   $\bm{\mbox{t}}$ and $\bm{\mbox{u}}$ as shown in eqs.
   (\ref{discreti}). Substrates and TGNR atoms are shown in grey and
   transparent yellow, respectively, while the set of carbon atoms
   corresponding to the TGNR centerline (adjacent line) are shown in
   red (black).}
\end{figure}

Every structure was optimized by an energy minimization method based
on gradient conjugate implemented in LAMMPS, with energy and force
tolerances of $10^{-8}$ and $10^{-8}$ eV/\AA, respectively. Thermal
fluctuations were simulated using Langevin thermostat, with timestep
set to 0.5 fs and thermostat damping factor set in 1 ps.

The set-up of the simulated experiments carried out here is shown in
Fig. \ref{fig2}. The nanoribbon chosen to investigate the SITWT
phenomenon is an initially straight hydrogen passivated armchair GNR
of about 600 \AA\mbox{} (33 \AA) length (width) to which a total
torsional strain of $4\pi$ (two full turns) was previously
applied. Its initial linking number is, then, $Lk = 2$. TGNRs in
Fig. \ref{fig2} were drawn in a transparent color in order to
facilitate the observation of the shape of their centerlines,
highlighted in red. The substrates are modeled as fixed graphene
layers.

Fig. \ref{fig2} shows two different configurations of suspended TGNRs
that have the same value of $Lk=2$ but different values of $Tw$ and
$Wr$ (see Table \ref{tab1}). They were obtained from two different
pathways as described below and will be considered for the experiment
of moving substrates. One of them came from bringing the extremities
of the TGNR into contact with two substrates followed by optimization.
The structure was, then, simulated for 4 and 8 ns at 300 and 1000 K,
respectively, in order to verify its thermal stability under the
suspended configuration. Optimization of these structures at the end
of the thermal simulations revealed no difference in their
corresponding configurations. Fig. \ref{fig2}b shows this optimized
structure.

Before proceeding to the dynamical simulations of the moving
substrates experiment, I have looked for other possible equilibrium
configurations of suspended TGNRs with the same $Lk=2$. The recent
work by Savin {\it et al.}~\cite{38}, then, came to my knowledge.
There, a particular TGNR, also having $Lk=2$, was fully laid on a
substrate and the final configuration displayed two {\it loop-like}
structures named by them as {\it twistons}. After testing the
formation of the same two {\it twistons}, I moved the structure to a
suspended configuration on two separate substrates, and simulated it
by 8 ns at 300 K. Then, the configuration shown in Fig. \ref{fig2}a
was found. Further simulation of this structure at 1000 K, made it to
become similar to that of Fig. \ref{fig2}b, indicating that they might
have similar cohesive energies. In fact, Table \ref{tab1} shows that
the optimized cohesive energies of the structures shown in
Figs. \ref{fig2}a and \ref{fig2}b are very close. The files containing
the coordinates of the atoms of the structures shown in
Fig. \ref{fig2} are provided in Supplemental Material~\cite{sm}.

\section{Results}
\label{sec3}

\subsection{Test of the numerical calculation of $Wr$ and $Tw$}

The centerline and its adjacent line of the TGNRs considered here
possess 285 carbon atoms. Therefore, the eqs. (\ref{x}) and (\ref{y})
possess 285 coordinates. Before using the discretization of
eqs. (\ref{tw}) and (\ref{wr}) to calculate $Tw$ and $Wr$ of the
TGNRs, as described and explained in the previous section, the
accuracy of the eqs. (\ref{discreti}) was tested with two discretized
special curves: i) a helical curve closed by straight segments similar
to that shown in Fig. 4 of Fuller's paper~\cite{41}, and ii) a
discretized almost straight TGNR, to which two turns were previously
applied (the same structure used to draw the panels (b) and (c) of
Fig. \ref{fig1}). According to Fuller, the {\it writhe} of a ribbon
having that particular centerline curve can be calculated by the
formula $Wr=n-n\sin\alpha$, where $\alpha$ is the helix pitch angle of
the helical part of the curve, and $n$ is the number of turns. I
generated a list of points following the helical curve with $n=2$,
radius = $1$ and pitch = $4\pi$, which, from the Fuller's formula,
provides $Wr = 0.585786$. Using the proposed discretization method,
the result for the numerical calculation of {\it writhe} of the
discretized curve i), with 285 points, is $Wr \simeq 0.5832$. The
second curve (in fact, two curves are needed, the centerline and an
adjacent line) was considered for the calculation of the total twist,
$Tw$, since the {\it writhe} of a straight curve is zero. $Tw$ of the
almost straight $4\pi$-twisted TGNR, whose centerline and adjacent
line also have 285 points, was obtained as $Tw \simeq
1.987$. Therefore, the estimated uncertainty in the calculations of
$Wr$ and $Tw$ using the present method is $\lesssim 0.02$. Wolfram
Mathematica scripts and the data points used to calculate $Tw$ and
$Wr$ of curves i) and ii) are provided in Supplemental
Material~\cite{sm}.

\subsection{$Wr$ and $Tw$ of static TGNRs}

Using the above discretization method, $Tw$ and $Wr$ of the structures
shown in Fig. \ref{fig2} were calculated. Table \ref{tab1} shows the
values of $Tw$, $Wr$ and the sum $Tw+Wr$ for these two TGNRs, showing
that although they have different values of $Tw$ and $Wr$, their sum
is $ \simeq 2$ within the uncertainties of the calculation method.
These results confirm the validity of the {\it linking number}
theorem, eq. (\ref{cwf}), and the SITWT. The possibility of performing
additional control of the $twist$ and $writhe$ of the TGNR, while
keeping $Lk$ conserved, and the results for the dynamical tests of the
SITWT will be shown in the next subsection.

\begin{table}[h!] 
 \caption{\label{tab1} Energy per atom, $E$ [eV/atom], $Wr$, $Tw$ and
   the sum $Wr+Tw$ corresponding to the TGNR structures shown in
   Fig. \ref{fig2}.}
 \begin{ruledtabular}
   \begin{tabular}{ccc}
     & Fig. \ref{fig2}a & Fig. \ref{fig2}b  \\
     \hline %\hline
     $E$ [eV/atom] & -7.0849 & -7.0848 \\
     $Wr$ & 0.323 & 0.457  \\
     $Tw$ & 1.663 & 1.524 \\
     $Wr+Tw$ & 1.986 & 1.981 
 \end{tabular}
 \end{ruledtabular}
\end{table}

The results shown in Table \ref{tab1} raise an important issue
regarding the determination of the total amount of twist of a given
TGNR. Although the TGNRs of Fig. \ref{fig2} initially received a
torsional strain of $4\pi$, as soon as the TGNR extremities touched
the substrates, its total amount of twist became no longer $4\pi$
anymore ($4\pi$ corresponds to $Tw=2$). Besides, although both
configurations shown in Fig. \ref{fig2} have $Lk=2$, both have neither
$Tw = 2$ nor the same $Tw$. $Tw$ calculated from eq. (\ref{tw})
represents the real values of the total twist of the nanoribbon. As
the electronic properties of GNRs depend on the amount of twist
applied to them~\cite{14,15,16,35,20,22}, it is important to know the
real value of the twist in order to correctly determine the
structure-property relationships in TGNRs. Subsection \ref{tuning}
shows an example of how to find out the right distance between the
substrates on which a TGNR of $Lk=2$ presents a chosen value of the
$Tw$.

\subsection{Dynamical interconvertion of $Wr$ and $Tw$ in TGNRs}

In view of the problem mentioned in the previous section and the need
for precise determination of the total twist of a TGNR, the present
experimental proposal of moving substrates with suspended TGNRs might
come in handy. The reason is that by simply controling the substrate
distance, the amount of twist of a TGNR can be determined. In order to
demonstrate that, I simulated several cycles of movements of the
substrates using the structures shown in Figs. \ref{fig2}a and
\ref{fig2}b. From these simulations, using the discretization method
described in Subsection \ref{discretizar}, $Tw$, $Wr$ and $Tw+Wr$ were
calculated as function of time. One cycle of the numerical experiment
consists of moving both substrates, one towards the other until almost
touching, so ``closing'' them, then inverting the velocities and
moving back to the initial distance, so ``opening'' them. Each
substrate was moved at an absolute speed of 0.2 \AA/ps, then the
effective approaching or going away speed was 0.4 \AA/ps. For an
initial maximum distance of $\sim$ 320 \AA, one cycle takes 1.6 ns. In
the numerical simulations of the experiment, the atoms of the TGNR of
Fig. \ref{fig2}a (Fig. \ref{fig2}b) were thermostated at 300 K (1000
K) in order to verify if conditions close to realistic situations
influence the results. The atoms of the substrates, however, were not
thermostated.

In order to calculate the dependence of $Wr$ and $Tw$ of the TGNRs
with time, one frame of the system was collected every 20 ps, or 50
frames were collected per nanosecond. Every frame provides the sets of
carbon atoms positions of the TGNR centerline and adjacent line and,
from them, the quantities given in eq. (\ref{discreti}) were
calculated. The summation of $Wr$ and $Tw$ allows for the verification
of the {\it linking number} theorem and, consequently, once more, the
legitimacy of the SITWT.

\begin{figure}[h!]
 \includegraphics[width=8.5cm,keepaspectratio]{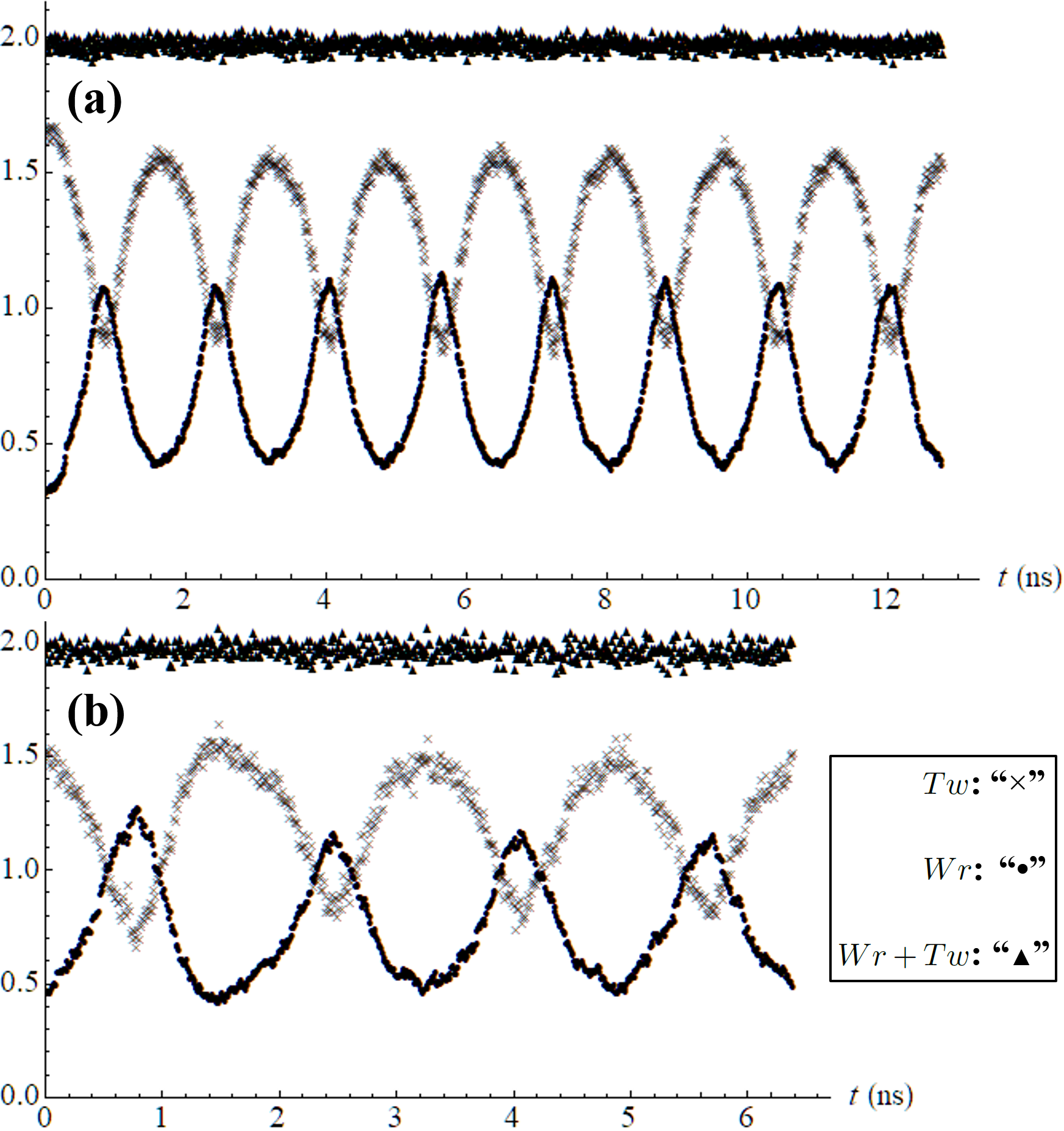}
 \caption{\label{fig3} Variation of {\it writhe}, $Wr$ (circles), {\it
     total twist}, $Tw$ (crosses), and the sum $Wr + Tw$ (triangles)
   with time for (a) the TGNR of Fig. \ref{fig2}a, 8 cycles simulated
   at 300 K and (b) the TGNR of Fig. \ref{fig2}b, 4 cycles simulated
   at 1000 K. }
\end{figure}

Fig. \ref{fig3}a (\ref{fig3}b) shows $Wr$, $Tw$ and $Wr + Tw$ as
function of time, during 8 (4) cycles of closing and opening the
substrates with the TGNRs of Fig. \ref{fig2}a (\ref{fig2}b) simulated
at 300 K (1000 K). Fig. \ref{fig3} shows that $Wr$ and $Tw$ oscillate
between minimum and maximum values during the cycles. The maximum
(minimum) value of $Wr$ happens for the substrates closed (opened) and
contrary for $Tw$. The rate of changing $Wr$ and $Tw$ is not uniform
despite the constancy of the speed of moving substrates. The rate
increases (decreases) when the substrates get closed (far) one to each
other, what suggests that longer the suspended TGNR easier to
fine-tune its total twist. Movies from S1 to S4 in Supplemental
Material~\cite{sm} show upper and lateral views of one cycle of the
experiment with both the TGNRs shown in Fig. \ref{fig2}. The movies
allow to see the change of the centerline as the substrates get closed
and go away. 

Fig. \ref{fig3}, then, demonstrates the possibility of controlling the
amount of total twist of a TGNR by just changing the distance between
the substrates on which its ends are laid. The results for $Wr + Tw$
along the time show that the {\it linking number} theorem,
eq. (\ref{cwf}), is satisfied within thermal fluctuations and
uncertainties that come from the discrete method of calculating $Wr$
and $Tw$.

\begin{figure}[h!]
 \includegraphics[width=8.5cm,keepaspectratio]{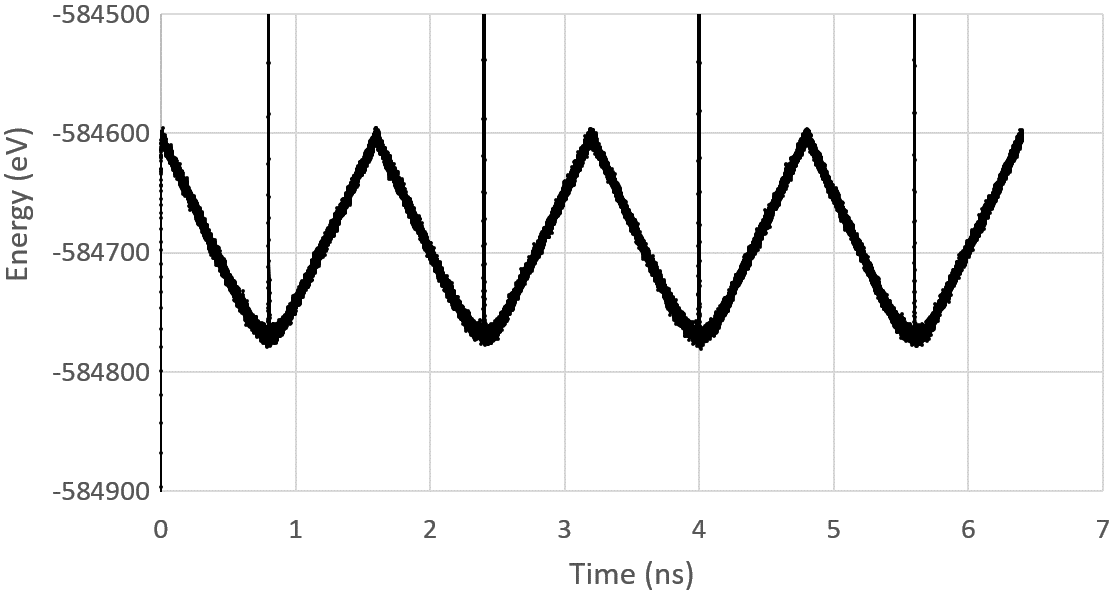}
 \caption{\label{fig4} Cohesive energy of the whole system composed by
   substrates $+$ TGNR of Fig. \ref{fig2}a as function of time, during
   4 cycles of the movement of the substrates.}
\end{figure}

Fig. \ref{fig4} displays the energy of the whole system during 4
cycles of movement of the substrates with the TGNR of Fig. \ref{fig2}a
simulated at 300 K. The energy decreases with the increase of the
contact between the TGNR and substrates (increased adhesion), and
back. The cusps in Fig. \ref{fig4} represent the subtle increase of
the energy of the system because the simulation allowed the substrates
to be almost in full contact. The rate of variation of the energy with
the time, calculated from the inclination of the curve in
Fig. \ref{fig4}, is $P\simeq41.3$ nW. It provides an estimate for the
external power necessary to carry out the SITWT process. Assuming the
force, $F$, needed to move the substrates is approximately constant,
using the equation $P=Fv$, with $v=0.4$ \AA/ps, it is found that
$F\simeq1$ nN. This value of force is within the range of actuation of
AFM microscopes~\cite{danielPaul1990}.

\subsection{Example of determination of the  distance between substrates
  to reach a chosen $Tw$}
\label{tuning}

From Fig. \ref{fig3} we see that the range of variation of the total
twist of the initially applied 2 turns (or 4$\pi$) TGNR is $0.8
\lesssim Tw \lesssim 1.6$. To ilustrate the possibility of chosing and
determining the total amount of twist of the TGNR, within
uncertainties of the method, let us find out the distance, $d$,
between the substrates, such that $Tw$ has the chosen value. Suppose
the desired value of the {\it total twist} of the TGNR is
$Tw=1$. Based on the present conditions of MD simulations,
\begin{equation}
  \label{dd}
   d = 320 - vt\, , \mbox{$d$ in \AA\mbox{ }and $t$ in ps,}
\end{equation}
where $v = 0.4$ \AA/ps is the simulated speed of approaching or moving
away the substrates. Taking the value of $t \approx 750$ ps, that
corresponds to $Tw \approx 1$ in Fig. \ref{fig3}a, we obtain $d \cong
20$ \AA.

Applications, other than controlling the electronic properties of the
TGNR, are the possibility of tunning thermal transport and mechanical
properties of TGNRs by fixing their amount of {\it twist}. As the {\it
  writhe} of a suspended TGNR varies with the distance between the
substrates in the present SITWT method, any physical property that
depends on ribbon shape can be also controlled by controlling the
substrate distance. These options expand the range of possible
applications of suspended TGNRs.

\section{Conclusions}
\label{sec4}

In summary, a method to adjust and determine the amount of twist of a
previously twisted GNR without the need of applying additional
rotation is presented and computationally demonstrated. The method
reveals the concept of a tension-free, ends-rotation-free, substrate
induced {\textit{twist to writhe transition}} in twisted
nanoribbons. The method relies on the adhesion forces between the
extremities of the twisted nanoribbon and the substrates, and on the
relatively low flexural rigidity of the ribbon. The {\it total twist},
$Tw$, {\it writhe}, $Wr$, and the sum $Tw+Wr$ were numerically
calculated for several configurations of suspended TGNRs, obtained
from MD simulations. In particular, the sum $Tw+Wr$ was compared to
the value of $Lk$ initially ascribed to the TGNR ($Lk=2$). The results
were shown to satisfy the {\it linking number} theorem,
eq. (\ref{cwf}), within the uncertainties of the methods and thermal
fluctuations. Estimates for the power and force needed to move the
substrates were presented based on the MD results. An application of
the method to the controlling of the total amount of twist of a TGNR
was also presented. The advantage of such a method is the possibility
of fine-tunning the total twist of a TGNR by simply moving the
substrates on which its extremities are laid. This method, then, might
be useful for experimentalists to manipulate TGNRs. It was also shown
that temperature does not affect the SITWT phenomenon, so the
experiment can be performed at different temperatures. I hope this
work motivates the development of new experiments and applications of
twisted nanoribbons.

\begin{acknowledgments}
I thank the Brazilian Agency CNPq for grant 311587/2018-6 and S\~{a}o
Paulo Research Foundation (FAPESP) for the grant \#2020/02044-9. This
research also used the computing resources and assistance of the John
David Rogers Computing Center (CCJDR) in the Institute of Physics
``Gleb Wataghin'', University of Campinas.
\end{acknowledgments}


\end{document}